\documentclass[11pt,reqno]{amsart}
\usepackage{amsmath,amssymb,mathrsfs,graphicx}
\usepackage[utf8x]{inputenc}
\usepackage[T1]{fontenc}
\usepackage[USenglish,english]{babel}
\usepackage{amsmath}
\usepackage{amsfonts}
\usepackage{latexsym}
\usepackage{cite}
\usepackage{float}
\usepackage{tikz}
\usepackage{pgfplots}
\usepackage{caption}
\usepackage{enumitem}
\usepackage{booktabs} 
\usepackage{subcaption}
\usepackage{mathtools}      
\mathtoolsset{showonlyrefs} 
\usepackage[export]{adjustbox}
\usepackage{amsthm}
\usepackage{amssymb,amscd}
\usepackage{mathtools}
\usepackage{xargs}
\usepackage{graphicx}
\usepackage{color}
\usepackage{enumitem}
\usepackage{multirow}
\usepackage{lmodern}
\usepackage{pdflscape}

\usepackage[text={7.3in, 9.6in}, centering]{geometry}
 
\usepackage{parskip}
\usepackage{microtype}

\usepackage{hyperref}

\newtheorem{thm}{Theorem}
\newtheorem*{thm*}{Theorem}

\usepackage{appendix}

\theoremstyle{definition}

\newcommand{\T}{\mathbf T}
\newcommand{\Tss}{\mathbf{T}_{ss}}

\newcommand{\tpq}{t_{pq}}
\newcommand{\spq}{s_{pq}}
\newcommand{\pp}{\mathbf{e}_1}
\newcommand{\qq}{\mathbf{e}_2}

\newcommand{\Xt}{\mathbf X_t}
\newcommand{\X}{\mathbf X}
\newcommand{\Xs}{\mathbf X_s}
\newcommand{\Xss}{\mathbf X_{ss}}
\newcommand{\Tt}{\mathbf T_t}
\newcommand{\RR}{\mathbb R}

\newcommand{\Hbf}{\mathbf H}

\title[]{Pseudorandomness of the Schr\"odinger map equation}
\author[S. Kumar]{Sandeep Kumar}
\address[S. Kumar]{Department of Quantitative methods, CUNEF Universidad, Madrid, Spain}
\email{sandeep.kumar@cunef.edu}

\date{\today}	
\begin{document}
\newenvironment{red}{\textcolor{red}}

\maketitle
\begin{abstract}
	
We present the random behaviour of the Schr\"odinger map equation, a geometric partial differential equation, by considering its evolution for regular polygonal curves in both Euclidean and hyperbolic spaces. The results obtained are consistent with those for the vortex filament equation, an equivalent form of the Schr\"odinger map equation in the Euclidean space, and thus, provide a novel extension to its usefulness as a pseudorandom number generator. 
 	
\end{abstract}

\section{Introduction}
Given a three-dimensional arc-length parametrized curve $\mathbf X$, and its tangent vector $\mathbf T=\Xs$, with arc-length parameter $s$, time $t$, their evolution can be described by
\begin{equation}
\label{eq:VFE}
\Xt = \Xs \wedge \Xss, \, \Tt = \T \wedge \Tss,
\end{equation}
where $\wedge$ is the cross-product. During the time evolution, $\T$ preserves its magnitude which allows us to assume its value on the unit sphere. The equation for $\T$ is called Schr\"odinger map onto the sphere and due to its geometric form, it can be expressed in different domains and images, while that for $\X$ is known as the vortex filament equation (VFE) \cite{Khesin,Ding1998}. The equation first appeared in 1906 as an approximation of a vortex filament under the Euler equations and since then has received a significant amount of attention, thanks to its rich geometric, yet simple form \cite{darios,arms}. We write $\wedge \equiv \wedge_{\pm}$, where the positive sign corresponds to the Euclidean space setting, i.e., $\T\in\mathbb{S}^2, \, \X\in\RR^3$, and the negative for the hyperbolic one, i.e., $\T\in\mathbb H^2, \X\in\mathbb{R}^{2,1}$. An important connection of \eqref{eq:VFE} with the cubic nonlinear Schr\"odinger equation was established in 1970, where the wave function $\psi$ describes the configuration of the curve through its curvature $\kappa$, torsion $\tau$, and solves \cite{hasimoto}
\begin{align}\label{eq:NLS-hyp}
	\psi_t = i\psi_{ss} \pm \frac{i}{2} \psi (|\psi|^2+A(t)), \ A(t)\in\RR, \ \psi(s,t)=\kappa(s,t) e^{\int_{ }^{s} \tau(s^\prime,t) ds^\prime}.
\end{align}

Besides the explicit solutions of \eqref{eq:VFE}--\eqref{eq:NLS-hyp} such as straight line, circle and helix, one important class is a one-parameter family of self-similar solution curves, that develop a corner in finite time \cite{GutierrezRivasVega2003}. The dynamics of \eqref{eq:VFE} for one-corner curves have been well-studied both theoretically and numerically and motivated the study of curves with multiple corners \cite{DelahozGarciaCerveraVega09,BanicaVega2016}. In particular, the evolution of regular planar curves in the Euclidean case capture some qualitative feature of real fluids; for instance, the axis-switching phenomenon observed in the case of non-circular jets \cite[Fig. 1.1]{GG,KumarPhd}. Later, it was shown that this evolution exhibits a pseudorandom behaviour, also a characteristic of real fluids \cite{HozVega2014b}. Recently, these results have been generalized to the cases of regular polygons with non-zero torsion and those in the Minkowski 3-space $\mathbb{R}^{2,1}$ \cite{HozKumarVega2019,HozKumarVega2020,Kumar2020,kumar2021recent}. These steps have successfully laid a base for the theoretical results of the multiple corner problem \cite{BanicaVega2018,BanicaVega2020}.  

In this article, our objective is to answer if the random behaviour mentioned above remains valid in the case of the regular polygons in the Minkowski 3-space and the helical polygons in the Euclidean space. In \cite{HozKumarVega2019, HozKumarVega2020, Kumar2020}, under uniqueness assumptions, the evolution of polygonal curves were described for rational times through an algebraic construction of the tangent vector $\T$, the solution curve $\X$, which was verified with strong numerical evidence. For a given rational time, we use this algebraic solution to define a complex number, whose real part is the triple product of three consecutive values of $\T$ and the imaginary part is the scalar product of two values of $\T$ \cite{HozVega2014b}. This form of complex numbers has a resemblance with some well-known pseudorandom generators such as explicit inversive congruential generators (EICG) \cite{eichenauer1993statistical,eichenauer1993explicit,HozVega2014b}. Furthermore, they can be calculated as a by-product of the evolution of \eqref{eq:VFE}, thus, without any additional computational cost! The structure of this article is the following. In Section \ref{sec:SMPforPolygons}, after a brief introduction to the regular polygon problem, for each of the regular polygon problems, we present three different cases according to the rational form of the time $t$. We derive relevant quantities and discuss their {\em randomness} in Section \ref{sec:Randomness}, followed by concluding remarks in Section \ref{sec:Conclusion}.

\section{Schr\"odinger map for polygonal curves}
\label{sec:SMPforPolygons}
In case of the evolution of \eqref{eq:VFE} for planar $l$-polygon as the initial datum, the corresponding initial data for \eqref{eq:NLS-hyp} translates as
$	\psi(s,0) = c_0 \sum_{k=-\infty}^{\infty} \delta(s-lk), $
with corners located at $s=lk$ and $c_0$ as in \cite[(26)]{HozKumarVega2020}. Then by using the Galilean invariance of \eqref{eq:NLS-hyp}  under uniqueness assumption, for a rational time $t=\tpq=(p/q)(l^2/(2\pi))$, $\gcd(p,q)=1$, 
$	\psi(s,\tpq) = ({l}/{\sqrt{q}}) \hat{\psi}(0,t) \sum_{k=-\infty}^{\infty}\sum_{m=0}^{q-1}G(-p, m,q) \delta\left(s-lk- {lm}/{q}\right),$
where the generalized quadratic Gau\ss\, sum can be written as
\begin{equation}
	G(-p,m,q)=
	\begin{cases}
		\sqrt{q}e^{i\theta_m}, \, & \text{if $q\equiv 1\bmod 2$}, \\
		\sqrt{2q}e^{i\theta_m}, \, & \text{if $q\equiv 0\bmod 2 \wedge q/2 \equiv m \bmod 2$}, \\
		0, \, & \text{if $q\equiv 0\bmod 2 \wedge q/2 \not\equiv m \bmod 2$}, 	
	\end{cases}
\end{equation}
for an angle $\theta_m$ dependent on $q$,  $\hat \psi(0,t)$ is a time-dependent constant. This expression implies that at time $\tpq$, depending on the denominator $q$, there will be $q \, (q/2)$ times more corners if $q$ is odd (even). Next, $\psi$ is modified by an appropriate scaling so that it takes the form
$	\Psi(s,\tpq) = \sum_{k=-\infty}^{k=\infty} \sum_{m=0}^{q-1} \rho_q e^{i\theta_m} \delta\left(s-lk- {lm}/{q}\right)$
where $\rho_q$, the angle between any two sides of the newly formed polygon, is calculated from the conservation law established in \cite{BanicaVega2018}. Using this, the generalized Frenet frame can be integrated for each $k, m$ and as a result, we obtain the rotation matrices $\Hbf_{k,m}$, which describes the transition across a corner and thus, the tangent vector $\T(\cdot, \tpq)$. Another integration with respect to $s$ yields the polygonal curve $\X(\cdot,\tpq)$ up to a rigid movement, which can be determined using the symmetries of the problems. However, in this article, we ignore this global rotation as the cross and scalar products are invariant under these transformations and hence, avoid complicated expressions.  

Similarly, the initial data for the helical polygon problem is given by 
$	\psi_\theta(s,0) = c_{\theta,0} e^{i\gamma s} \sum_{k=-\infty}^{\infty} \delta(s-rk), $
\cite{Kumar2020}; for example, $\gamma=\theta_0/l$ $r=l$, for the hyperbolic helical polygon (HHP), and $\gamma=M\theta_0/2\pi$, $r=2\pi/M$, for the circular helical polygon (CHP) and Euclidean helical polygon. Here, $\theta_0$ is the torsion angle, $M$, the number of sides, $c_{\theta,0}, \, l>0$ (see \cite{HozKumarVega2019,HozKumarVega2020,Kumar2020} for their expressions). By following a procedure similar to the one above, the algebraic solution can be constructed up to a rigid movement, whereas the numerical treatment for these problems is different (see \cite{HozKumarVega2019,HozKumarVega2020,Kumar2020} for more details). Below, we consider each of these problems and discuss them for three different cases according to the parity of $q$ and $q/2$. In this article, with some abuse of notation, throughout all the subsections, we use $\Psi$, as defined above.   
\subsection{Planar polygons in Minkowski 3-space}
\subsubsection{When $q\equiv 1\bmod 2$}
When $q$ is odd, 
$\Psi(s,\tpq) = \rho_q \sum_{m=0}^{q-1} e^{i\theta_m} \delta\left(s- {lm}/{q}\right), \ s\in[0, l),$
which implies that the vertices $\X_m$ are located at $s= {lm}/{q}$. Motivated by the ideas in \cite{HozVega2014b}, we calculate the triple product of $\T({lm^-}/{q})$, $\T({lm^+}/{q})\equiv\T({l(m+1)^-}/{q})$, $\T({l(m+1)^+}/{q})$, and the scalar product of $\T({lm^-}/{q})$, $\T({l(m+1)^+}/{q})$, and denote them by $x_{q,m}$ and $y_{q,m}$, respectively. We calculate the tangent vectors using the rotation matrix $\Hbf_m$ ($\equiv \Hbf_{0,m}$), by assuming that the tangent vector $\T({lm^-}/{q})=(1,0,0)$, and the generalized normal and binormal vectors, $\pp({{lm^-}/{q}})=(0,1,0)$, $\qq({{lm^-}/{q}})=(0,0,1)$. Hence, $\T({lm^+}/{q})$ is the first row of $\mathbf{H}_m$, and $\T({l(m+1)^+}/{q})$ is the first row of $\mathbf{H}_{m+1}\cdot\mathbf{H}_m$. As a result, we get the first quantity
\begin{align}
x_{q,m}&=\left(\T\left({lm^-}/{q}\right) \wedge_{-} \T\left({lm^+}/{q} \right)  \right) \circ_{-} 
\T\left({l(m+1)^+}/{q}\right) 
= \tilde{s}_{\rho_q}^2 s_{\Delta_m} = \tilde{s}_{\rho_q}^2 s_{\theta_{m+1}-\theta_m} =  \tilde{s}_{\rho_q}^2 \Im\left(e^{i\theta_{m+1}} e^{-i\theta_m}\right) \nonumber \\
&=\tilde{s}_{\rho_q}^2 \Im\left[{G(-p, m+1, q)} {\bar{G}(-p, m, q)}/{{q}} \right] =\tilde{s}_{\rho_q}^2 \Im\left[ e^{2\pi i \phi(p)(m+1)^2/q} e^{-2\pi i \phi(p)m^2/q} \right]  = \tilde{s}_{\rho_q}^2 \sin\left(\vartheta_{q,m}(p)\right)
\end{align}
where in the last step we have used the properties of quadratic Gauss sum (in particular, \cite[(29)]{HozVega2014b}), and $\phi(p)$ is the inverse of $p$ in the finite ring $\mathbb{Z}_q={0,1,\ldots, q-1}$, $\vartheta_{q,m}(p)={2\pi \phi(p)(2m+1)}/{q}$, $\tilde s_{\rho_q} = \sinh(\rho_q)$, $\tilde c_{\rho_q} = \cosh(\rho_q)$. The other quantity is the Minkowski scalar product of $\T({lm^-}/{q})$, and $\T({l(m+1)^+}/{q})$:
\begin{align}
	y_{q,m}&=\T\left({lm^-}/{q}\right) \circ_{-} \T\left({l(m+1)^+}/{q}\right) = \tilde{c}_{\rho_q}^2 + \tilde{s}_{\rho_q}^2 c_{\Delta_m} =\tilde{c}_{\rho_q}^2 + \tilde{s}_{\rho_q}^2 c_{\theta_{m+1}-\theta_m} 	
 = \tilde{c}_{\rho_q}^2 + \tilde{s}_{\rho_q}^2 \cos\left(\vartheta_{q,m}(p)\right).
\end{align}
Next, we construct a complex number with a $x_{q,m}$, and $y_{q,m}$ as the real and imaginary part, respectively,
\begin{align}
	z_{q,m}(p) = x_{q,m}+iy_{q,m}=i\tilde{c}_{\rho_q}^2 + i \tilde{s}_{\rho_q}^2 \exp\left(-i\vartheta_{q,m}(p)\right).
\end{align}
We note that for a fixed $q$, and $m$, $z_{q,m}$ depend only on $\phi(p)$, i.e., the inverse of $4p$ modulo $q$.

\subsubsection{When $q\equiv 2\bmod 2$}
When $q/2$ is odd, 
$	\Psi(s,\tpq) = \rho_q \sum_{m=0}^{q/2-1} e^{i\theta_{2m+1}} \ \cdot \delta\left(s - {l(2m+1)}/{q}\right)$,  $s\in[0, l),$
and the vertices $\X_{2m+1}$ are located at $s={l(2m+1)}/{q}$, and we calculate the triple product of $\T({l(2m-1)^-}/{q})$, $\T({l(2m-1)^+}/{q})\equiv\T({l(2m+1)^+}/{q})$, and the scalar product of $\T({l(2m-1)^-}/{q})$, and $\T({l(2m+1)^+}/{q})$. The computations are similar as those for the case when $q$ is odd, except a certain change in the subscripts, i.e., we replace $c_m$, $s_m$, by $c_{2m-1}$, $s_{2m-1}$, respectively, and $\Delta_m=\theta_{2m+1}-\theta_{2m-1}$:
\begin{align}
x_{q,m}
&=   \tilde{s}_{\rho_q}^2 s_{\theta_{2m+1}-\theta_{2m-1}} 
=\tilde{s}_{\rho_q}^2 \Im\left[{G(-p, 2m+1, q)} {\bar{G}(-p, 2m-1, q)}/{{q}} \right] \\
&= \tilde{s}_{\rho_q}^2 \sin\left({32\pi \phi_1(p)m}/{q}\right)
= \tilde{s}_{\rho_q}^2 \sin\left({2\pi \phi(p)m}/{(q/2)}\right)
= \tilde{s}_{\rho_q}^2 \sin\left(\vartheta_{q,m}(p)\right),
\end{align}
where $\vartheta_{q,m}(p)={2\pi \phi(p)m}/{(q/2)}$, and $\phi_1(p)$ is the inverse of $8p$ in $\mathbb{Z}_{q/2}$, i.e., $(8p) \phi_1(p)\equiv1\bmod(q/2)$, which implies that $8\phi_1(p)$, is the inverse of $p$ in $\mathbb{Z}_{q/2}$, which we call $\phi(p)$. Similarly, we obtain the second quantity,  
\begin{align}
	y_{q,m}=\T\left({l(2m-1)^-}/{q}\right) \circ_{-} \T\left({l(2m+1)^+}/{q}\right) &= 
	\tilde{c}_{\rho_q}^2 + \tilde{s}_{\rho_q}^2 \cos\left(\vartheta_{q,m}(p)\right).
\end{align}
Consequently, the complex number is
\begin{align}
	z_{q,m}(p)\equiv x_{q,m}+iy_{q,m}=i\tilde{c}_{\rho_q}^2 + i \tilde{s}_{\rho_q}^2 \exp\left(-i\vartheta_{q,m}(p)\right).
\end{align}
Thus, for a fixed $q$, and $m$, ${z}_{m,p}$ depend only on $\phi(p)$, i.e., the inverse of $p$ modulo $q/2$.
\subsubsection{When $q\equiv 0\bmod 2$}
When $q/2$ is even, $\psi(s,\tpq)$ can be expressed as
$	\Psi(s,\tpq) = \rho_q \sum_{m=0}^{q/2-1} e^{i\theta_{2m}} \delta\left(s- {2lm}/{q}\right)$,  $s\in[0, l),$
and the vertices $\X_{2m}$ are located at $s= {2lm}/{q}$. We calculate the triple product of $\T({2lm^-}/{q})$, $\T({2lm^+}/{q})\equiv\T({l(2m+2)^+}/{q})$, and the scalar product of $\T({2lm^-}/{q})$, and $\T({l(2m+2)^+}/{q})$. After a similar computation and a change of subscripts, i.e., replacing $c_m$, $s_m$, by $c_{2m}$, $s_{2m}$, respectively, and $\Delta_m=\theta_{2m+2}-\theta_{2m}$, we get
\begin{align}
	\label{eq:quant1q2even}
x_{q,m} 	&=  \tilde{s}_{\rho_q}^2 \Im\left(e^{i\theta_{2m+2}} e^{-i\theta_{2m}}\right) =\tilde{s}_{\rho_q}^2 \Im\left[{G(-p, 2m+2, q)} {\bar{G}(-p, 2m, q)}/{{q}} \right]= \tilde{s}_{\rho_q}^2 \sin\left(\vartheta_{q,m}(p)\right), \nonumber \\
	y_{q,m} &= 
	\tilde{c}_{\rho_q}^2 + \tilde{s}_{\rho_q}^2 \cos\left(\vartheta_{q,m}(p)\right), \, \text{and} \,  
	z_{q,m}(p)\equiv i\tilde{c}_{\rho_q}^2 + i \tilde{s}_{\rho_q}^2 \exp\left(-i\vartheta_{q,m}(p)\right),
\end{align}
where $\vartheta_{q,m}(p)={2\pi \phi(p)(2m+1)}/{q}$. We note that for a fixed $q$, and $m$, ${z}_{m,p}$ depend only on $\phi(p)$, i.e., the inverse of $p$ modulo $q$.

\subsection{Helical polygons in Minkowski 3-space} Depending on the axis of rotation, there are two types of helical curves in the Minkowski 3-space \cite{Kumar2020}. We consider the hyperbolic helical polygon and for brevity, omit the circular helical polygon as the calculations are similar.
\subsubsection{When $q\equiv 1\bmod 2$ (HHP)}
When $q$ is odd, 
$	\Psi(s,\tpq) = \rho_q \sum_{m=0}^{q-1} e^{i(\xi_m+m\theta_0/q)} \-\delta\left(s-{l\theta_0 p}/{(\pi q)}- {lm}/{q}\right), \ s\in[0, l),$
which implies that the vertices $\X_m$ are located at $s= {lm}/{q}+ {l\theta_0 p}/{(\pi q)}= {lm}/{q}+\spq$. We calculate the triple product of $\T\left(\left({lm}/{q}+\spq\right)^-\right)$,
$\T\left(\left({lm}/{q}+\spq\right)^+\right)\equiv \T\left(\left({l(m+1)}/{q}\-+\spq\right)^-\right)$,
$\T\left(\left({l(m+1)}/{q}+\spq\right)^+\right)$,
and the scalar product of $\T\left(\left({lm}/{q}+\spq\right)^-\right)$, and $\T\left(\left({l(m+1)}/{q}+\spq\right)^+\right)$, and by denoting them by $x_{m,q}$ and $y_{m,q}$, respectively, with $\zeta_m=\xi_m+m\theta_0/q$, $\Delta_m = \zeta_{m+1}-\zeta_m$,  we have
\begin{align}
	\label{eq:quant1qoddHHP}
	x_{q,m} &= \tilde{s}_{\rho_q}^2 s_{\Delta_m} = \tilde{s}_{\rho_q}^2 s_{\zeta_{m+1}-\zeta_m} =  \tilde{s}_{\rho_q}^2 \Im\left(e^{i\xi_{m+1}} e^{-i\xi_m} e^{i\theta_0/q} \right) = \tilde{s}_{\rho_q}^2 \sin\left(\vartheta_{q,m}^\theta(p)\right), \nonumber \\
	y_{q,m} &= \tilde{c}_{\rho_q}^2 + \tilde{s}_{\rho_q}^2 c_{\Delta_m} =\tilde{c}_{\rho_q}^2 + \tilde{s}_{\rho_q}^2 c_{\zeta_{m+1}-\zeta_m} = \tilde{c}_{\rho_q}^2 + \tilde{s}_{\rho_q}^2 \cos\left(\vartheta_{q,m}^\theta(p) \right), \nonumber \\	
	{z}_{q,m}(p)&\equiv x_{q,m} + i y_{q,m}=i\tilde{c}_{\rho_q}^2 + i \tilde{s}_{\rho_q}^2 \exp\left(- i \vartheta_{q,m}^\theta(p) \right),
\end{align}
where $\vartheta_{q,m}^\theta=({2\pi \phi(p)(2m+1)+\theta_0)}/{q}$, and for a fixed $q$, and $m$, ${z}_{m,p}$ depend only on $\phi(p)$, i.e., the inverse of $4p$ modulo $q$. 
\subsubsection{When $q\equiv 2\bmod 4$ (HHP)}
When $q/2$ is odd, 
	$\Psi(s,\tpq) = \rho_q \sum_{m=0}^{q/2-1} e^{i(\xi_{2m+1}+(2m+1)\theta_0/q)}\delta\left(s- \spq - {l(2m+1)}/{q}\right)$, $s\in[0, l),$ 
and the vertices $\X_{2m+1}$ are located at $s= {l(2m+1)}/{q}+ \spq$, and we calculate the complex number as 
\begin{align}
	\label{eq:quant1q2oddHHP}
	x_{q,m}= \tilde{s}_{\rho_q}^2 \sin\left(\vartheta_{q,m}^\theta(p)\right), \, y_{q,m}= \tilde{c}_{\rho_q}^2 + \tilde{s}_{\rho_q}^2 \cos\left(\vartheta_{q,m}^\theta(p)\right), 
	 {z}_{q,m}(p)=i\tilde{c}_{\rho_q}^2 + i \tilde{s}_{\rho_q}^2 \exp\left(-i\vartheta_{q,m}^\theta(p)\right),
\end{align}
where $\vartheta_{q,m}^\theta(p)=({2\pi \phi(p)m+\theta_0})/{(q/2)}$, and for a fixed $q$, and $m$, $ {z}_{m,p}$ depend only on $\phi(p)$, i.e., the inverse of $p$ modulo $q/2$.
\subsubsection{When $q\equiv 0\bmod 2$}
When $q/2$ is even, 
$	\Psi(s,\tpq) = \rho_q \sum_{m=0}^{q/2-1} e^{i(\xi_{2m}+2m\theta_0/q)} \delta\left(s- \spq - {2lm}/{q}\right)$,  $s\in[0, l),$
and the vertices $\X_{2m}$ are located at $s={2lm}/{q}+\spq $, and we calculate
\begin{align}
	\label{eq:quant1q2evenHHP}
x_{m,q}&= \tilde{s}_{\rho_q}^2 \sin\left(\vartheta_{q,m}^\theta(p) \right), \, y_{m,q}= \tilde{c}_{\rho_q}^2 + \tilde{s}_{\rho_q}^2 \cos\left( \vartheta_{q,m}^\theta(p) \right), \, {z}_{q,m}(p) =i\tilde{c}_{\rho_q}^2 + i \tilde{s}_{\rho_q}^2 \exp\left(-i \vartheta_{q,m}^\theta(p) \right),
\end{align}
where $\vartheta_{q,m}^\theta(p)={(2\pi \phi(p)(2m+1)+2\theta_0)}/{q}$, and for a fixed $q$, and $m$, ${z}_{m,p}$ depend only on $\phi(p)$, i.e., the inverse of $p$ modulo $q$.

\subsection{Helical polygons in Euclidean space}
\subsubsection{When $q\equiv 1\bmod 2$}
When $q$ is odd, 
$\Psi(s,\tpq) = \rho_q \sum_{m=0}^{q-1} e^{i(\xi_m+m\theta_0/q)} \delta\left(s- {2\theta_0p}/{(Mq)}- {2\pi m}/{(Mq)}\right), \ s\in\left[0, {2\pi}/{M}\right),$
which implies that the vertices $\X_m$ are located at $s={2\pi m}/{(Mq)}+\spq$, and $\spq={2\theta_0p}/{(Mq)}$ \cite{HozKumarVega2019}. The tangent vector has been calculated using the rotation matrix given by \cite[(18)]{HozVega2014b} and $\pp$, $\qq$, are chosen the same as before. The quantities resulting from the triple product and scalar product as defined in \cite{HozVega2014b}, yield 
\begin{align}
	x_{q,m}&= s_\rho^2 \sin\left(\vartheta_{q,m}^\theta(p)\right), \, y_{q,m}= c_\rho^2 - s_\rho^2 \cos\left(\vartheta_{q,m}^\theta(p)\right), \,\nonumber\\	
	z_{q,m}(p) &\equiv x_{q,m}+i y_{q,m} =i c_\rho^2 - i s_\rho^2 \exp\left(i\vartheta_{q,m}^\theta(p)\right),  
\end{align}	
where $\vartheta_{q,m}^\theta(p)=({2\pi \phi(p)(2m+1)+\theta_0})/{q}$, $s_{\rho_q}=\sin(\rho_q)$, $c_{\rho_q}=\cos(\rho_q)$, and for a given $\theta_0$ and fixed $q$, $z_{q,m}(p)$ depends on the the inverse of $4p$ modulo $q$.  
\subsubsection{When $q\equiv 2\bmod 4$}
In this case, 
$\Psi(s,\tpq) = \rho_q \sum_{m=0}^{q-1} e^{i(\xi_{2m+1}+(2m+1)\theta_0/q)} \delta\left(s-\spq -{2\pi (2m+1)}/{(Mq)}\right)$, $s\in\left[0, {2\pi}/{M}\right),$ and the vertices $\X_m$ are located at $s= {2\pi (2m+1)}/{Mq}+\spq$, and after computations, we get 
\begin{align}
	x_{q,m}= s_{\rho_q}^2 \sin\left(\vartheta_{q,m}^\theta(p)\right), \, 
	y_{q,m}= c_{\rho_q}^2 - s_{\rho_q}^2 \cos\left(\vartheta_{q,m}^\theta(p)\right), \,
	z_{q,m}(p) =i c_{\rho_q}^2 - i s_{\rho_q}^2 \exp\left(i \vartheta_{q,m}^\theta(p) \right),
\end{align}
where $\vartheta_{q,m}^\theta(p)={(2\pi \phi(p)m+\theta_0)}/{(q/2)}$,  and for a given $\theta_0$, for a fixed $q$, and $m$, ${z}_{m,p}$ depend only the inverse of $p$ modulo $q/2$. 
\subsubsection{When $q\equiv 0\bmod 2$}
When $q/2$ is even, 
$\Psi(s,\tpq) = \rho_q \sum_{m=0}^{q-1} e^{i(\xi_{2m}+(2m)\theta_0/q)} \delta\left(s- \spq - {4\pi m}/{(Mq)}\right)$,  $s\in\left[0, {2\pi}/{M}\right),$
and the vertices $\X_m$ are located at $s={4\pi m}/{(Mq)}+\spq$, after computations similar as before,  
\begin{align}
	x_{q,m}= s_{\rho_q}^2 \sin\left(\vartheta_{q,m}^\theta(p)\right), \, y_{q,m}= c_{\rho_q}^2 - s_{\rho_q}^2 \cos\left(\vartheta_{q,m}^\theta(p)\right), \, z_{q,m}(p) =i c_{\rho_q}^2 - i s_{\rho_q}^2 \exp\left(i\vartheta_{q,m}^\theta(p)\right).
\end{align}
where $\vartheta_{q,m}^\theta(p)=({2\pi \phi(p)(2m+1)+2\theta_0})/{q}$, and for a given $\theta_0$, and fixed $q$, and $m$, ${z}_{m,p}$ depend only on $\phi(p)$, i.e., the inverse of $p$ modulo $q$.

These calculations amount us to conclude the following theorem:

\begin{thm}
\label{thm:mainresult}
	Given the the triple product of three consecutive tangent vectors and the scalar product of a tangent vector and the second next one, calculated for each of the four cases of the polygonal curves using respective definitions, these quantities depend on $\phi(p)$.
 	Moreover, the complex number with the real part as the triple product and imaginary part as the scalar product mentioned above, is given by
\begin{equation}
 			\label{eq:z_lpolygon}
 			 z_{q,m}(p) =
 			\begin{cases}
 				i\tilde{c}_{\rho_q}^2 + i \tilde{s}_{\rho_q}^2 \exp\left(-i\vartheta_{q,m}(p)\right), & \text{planar $l$-polygon},  \\
 				i\tilde{c}_{\rho_q}^2 + i \tilde{s}_{\rho_q}^2 \exp\left(-i\vartheta_{q,m}^\theta(p)\right), & \text{Hyperbolic and Circular helical polygons},  \\
 				i {c}_{\rho_q}^2 - i {s}_{\rho_q}^2 \exp\left(i\vartheta_{q,m}^\theta(p)\right), & \text{Euclidean helical polygon},  
 				\end{cases}
 		\end{equation}
 	where  		
	 	\begin{equation}
	 		\label{eq:z_eucheli}
			\vartheta_{q,m}^\theta(p) =
	 		\begin{cases}
			(2\pi \phi(p)(2m+1)+\theta_0)/q, \, \, \text{where} \, \, \phi(p)=(4p)^{-1} \bmod q, & \text{if} \, \, q \equiv 1 \bmod 4, \\
				 			
			(2\pi \phi(p)m+\theta_0)/(q/2), \, \, \text{where} \, \,\phi(p)=p^{-1} \bmod (q/2), & \text{if} \, \, q \equiv 2 \bmod 4, \\
			
			(2\pi \phi(p)m+2\theta_0)/q, \, \, \text{where} \, \,\phi(p)=p^{-1} \bmod q, & \text{if} \, \, q \equiv 0 \bmod 4, \\
	 		\end{cases} 		
	 	\end{equation}
\end{thm}
and $\vartheta_{q,m}(p)=\vartheta_{q,m}^\theta(p)$, with $\theta_0=0$.
\section{Randomness}
\label{sec:Randomness}
An important outcome of Theorem \ref{thm:mainresult} is that the calculation of the complex number in each of the four cases reduces to finding the inverses in finite rings, i.e., calculation of $\phi(p)$. The existence and uniqueness of $\phi(p)$ follows from the fact that $p$ and $q$ are coprime, and thus, in the finite ring corresponding to a given $q$, there will be $\varphi(q)$ different values of $p$ and its inverse \cite{HozVega2014b}. Here, $\varphi(q)$ is Euler's totient function which counts the number of positive integers up to $q$ that are coprime to $q$. For a given $\theta_0\geq 0$, the form of the complex number in each case indicates that the maximum number of different values is obtained when $\gcd(q, 2m+1)=1$,
for $q\not\equiv2\bmod 4$ and $\gcd(q/2,m)=1$, for $q\equiv2\bmod 4$. Without any loss in generality, we assume $m=0$, $m=1$, when $q\not\equiv2\bmod 4$, and $q\equiv2\bmod 4$, respectively. Consequently, denoting the resulting quantities by $z_q(p)$, there are $\varphi(q)$ different complex numbers present on the same circumference whose calculation is immediate through several well-known algorithms, e.g., the extended Euclidean algorithm. Thus, for a given problem and $q$, the dependence of $z_q(p)$ on $\phi(p)$ hint that the random behaviour of latter, can induce that in $z_q(p)$, as well \cite{HozVega2014b}. 

 
A classical choice for the generating pseudorandom numbers is the linear congruential generator (LCG) where for a given $q\in\mathbb{N}$ (large enough) and $a, b, x_0 \in \mathbb{Z}$, 
$	x_{n+1} \equiv a x_n + b \bmod q, \, n\geq 0,$
is a linear congruential sequence of nonnegative integers smaller than $q$. The role of $q, a, b , x_0$, is crucial and with a normalization $x_n/q = u_n$, $n\geq 0$, a uniformly distributed sequence $(u_n)_{n\geq 0} \in [0,1)$, of linear congruential pseudorandom numbers is obtained\cite{knuth2014art}. However, LCGs suffer from poor uniformity, short periods, and susceptibility to correlation, and to tackle that, nonlinear random generators are introduced \cite{entacher1998bad}. In particular, the inversive congruential generators (ICGs) leverage the concept of modular inversion to provide enhanced randomness, longer periods, and reduced correlation between generated numbers \cite{eichenauer1988marsaglia}. These are often expressed as 
$	x_{n+1} \equiv a \bar x_n+b \bmod q, \, n\geq 0,$
with $q$ prime, $a\not\equiv0\bmod q$, where $\bar x_n$ is the multiplicative inverse of $x_n$. In particular, the explicit inversive congruential generators (EICGs) were found useful for the regular planar polygon problem \cite{HozVega2014b}. For $q$ prime, $a\not\equiv 0 \bmod q$, these can be expressed as
\begin{equation}
	\label{eq:EICG}
	x_n\equiv \overline{an+b} \bmod q, \, n\geq 0,
\end{equation}
i.e., $\{x_0, x_1, \ldots, x_{q-1}\}=\mathbb{Z}_q$. The normalization $u_n = x_n/q$ ensures the uniformity test for equidistribution in $[0,1)$, while the statistical independence is studied through a serial test. This requires studying the discrepancy of $k$-dimensional tuples of pseudorandom numbers where $k\geq 2$  \cite{eichenauer1993statistical}. We refer the reader to \cite{HozVega2014b} where a detailed discussion of these properties is presented along with the relevant results. Moreover, different choice of $q$, such as $q$, $q/2$ prime and $q=2^\omega$ are considered. With that, it can be concluded that EICGs possess very good structural and statistical independence properties, and are very similar to the one we discuss in this work. For instance, for a given $q$, 
\begin{equation}
	z_{q}(p) =
	\begin{cases}
 			 i\tilde{c}_{\rho_q}^2 + i \tilde{s}_{\rho_q}^2 \exp\left(-2\pi i u_p\right), \, & \text{planar $l$-polygon},\\ 
			 i\tilde c_{\rho_q}^2 + i \tilde s_{\rho_q}^2 \exp\left(-2\pi i u_p-\frac{i \theta_0}{q}\right),  \, & \text{Hyperbolic and Circular helical polygons}, \\
	 		 ic_{\rho_q}^2 - i s_{\rho_q}^2 \exp\left(2\pi i u_p+\frac{i \theta_0}{q}\right),
	 		 \, & \text{Euclidean helical polygon},
	\end{cases}
\end{equation}
where from  \eqref{eq:EICG}
\begin{equation}
	\label{eq:z_eicgform}
	u_p = 
	\begin{cases}
		 x_p/q, \, x_p\equiv \overline{4p} \bmod q, \text{i.e.}, \, a\equiv 4\bmod q, \, b\equiv 0\bmod q, & \text{if $q$ is odd prime}, 		\\
		 x_p/(q/2), \, x_p\equiv \overline{p} \bmod q/2, \text{i.e.}, \,  a\equiv 1\bmod q/2, \, b\equiv 0\bmod q/2, & \text{if $q/2$ is prime},	\\
		 x_p/q, \, x_p\equiv \overline{2p-1} \bmod q,  \text{i.e.},\,  a\equiv 1\bmod q, \, b\equiv -1\bmod q, & \text{if $q=2^\omega$}.		 
	\end{cases}
\end{equation}
Following the results in \cite{eichenauer1993statistical}, $u_p$ is a sequence of pseudorandom numbers uniformly distributed in the interval $(0, 1)$ \cite{HozVega2014b}. Through Theorem \ref{thm:mainresult} these sequences of pseudorandom numbers are uniformly distributed on the circumference with center-radius pair ($i\tilde c_{\rho_q}^2, \tilde s_{\rho_q}^2)$, and ($i c_{\rho_q}^2, s_{\rho_q}^2)$, for different expression of $\rho_q$ \cite{HozKumarVega2019,HozKumarVega2020,Kumar2020}. Note that these numbers are distributed in the interval $(0, 1)$, instead of $[0,1)$, since the case $p=0$, has to be omitted in the first two cases, but not in the third case. Finally, it's worth mentioning the compound approach in \cite{eichenauer1993explicit}, where the existing form $z_q(p)$ for different primes $q_1, q_2, \ldots, q_n\geq 5$, for any $n$, can be combined to to form different expressions that enjoy good random properties \cite{HozVega2014b}.

\section{Conclusion}
\label{sec:Conclusion}
In this article, we study the evolution of \eqref{eq:VFE} for initial data as planar and non-planar polygons in the Minkowski 3-space and the latter in Euclidean space. By calculating the triple product and the scalar product of the tangent vectors and taking them as the real and imaginary parts of a complex number, we propose four different candidates for generating pseudorandom numbers located uniformly on a circumference.  These findings show that the dynamics of \eqref{eq:VFE} are consistent with that in the Euclidean planar polygon case and also provide new classes of pseudorandom number generators. It is important to note that although a crude approximation of Euler equations, \eqref{eq:VFE} exhibit several interesting features of real fluids when solved for polygonal initial data. First, the axis-switching phenomenon mentioned earlier in the introduction and in this work the random behaviour, which is also a key property of several natural phenomena. It is in addition to the multifractality observed in $\X(0,t)$, i.e., the trajectory of a particle located on the polygonal curve \cite{HozVega2014,HozKumarVega2019,HozKumarVega2020}. Hence, this unusual (random) behaviour resulting from a differential equation (a deterministic model), not only makes its role stronger in describing the physical phenomenon of vortex filaments but also unveils a new feature of its geometrical form, which remains to be explored to its fullest. 

\bibliography{references}
\bibliographystyle{ieeetr}

\end{document}